\documentclass[aps,twocolumn,groupedaddress,showpacs]{revtex4}

\usepackage{amssymb}
\usepackage{amsmath}
\usepackage{epsf,latexsym}
\usepackage{times}

\newenvironment{proof}[1][Proof]{\noindent\textbf{#1.} }{\ \rule{0.5em}{0.5em}}

\newcommand{\ket}[1]{| #1 \rangle}

\newcommand{\be}{\begin{equation}}
\newcommand{\ee}{\end{equation}}
\newcommand{\bae}{\begin{eqnarray*}}
\newcommand{\eae}{\end{eqnarray*}}
\def\CC{{\rm\kern.24em \vrule width.04em height1.46ex depth-.07ex
    \kern-.30em C}}
\def\P{{\rm I\kern-.25em P}}

\def\bbbc{{\mathchoice {\setbox0=\hbox{$\displaystyle\rm C$}\hbox{\hbox
to0pt{\kern0.4\wd0\vrule height0.9\ht0\hss}\box0}}
{\setbox0=\hbox{$\textstyle\rm C$}\hbox{\hbox
to0pt{\kern0.4\wd0\vrule height0.9\ht0\hss}\box0}}
{\setbox0=\hbox{$\scriptstyle\rm C$}\hbox{\hbox
to0pt{\kern0.4\wd0\vrule height0.9\ht0\hss}\box0}}
{\setbox0=\hbox{$\scriptscriptstyle\rm C$}\hbox{\hbox
to0pt{\kern0.4\wd0\vrule height0.9\ht0\hss}\box0}}}}
\def\bbbz{{\mathchoice {\hbox{$\sf\textstyle Z\kern-0.4em Z$}}
{\hbox{$\sf\textstyle Z\kern-0.4em Z$}}
{\hbox{$\sf\scriptstyle Z\kern-0.3em Z$}}
{\hbox{$\sf\scriptscriptstyle Z\kern-0.2em Z$}}}}

\begin{document}

\title{Purity and State Fidelity of Quantum Channels via Hamiltonians}
\author{Paolo Zanardi}
\affiliation{Institute for Scientific Interchange (ISI), Villa Gualino, Viale
Settimio Severo 65, I-10133 Torino, Italy}
\affiliation{Department of Mechanical Engineering, Massachusetts Institute of
Technology, Cambridge Massachusetts 02139}
\author{Daniel A. Lidar}
\affiliation{Chemical Physics Theory Group, Chemistry Department, University of Toronto,
80 St. George Street, Toronto, Ontario M5S 3H6, Canada}

\begin{abstract}
We associate to every quantum channel $T$ acting on a Hilbert space
$\mathcal{H}$ a pair of Hamiltonian operators over the symmetric
subspace of $\mathcal{H}^{\otimes \,2}$. The expectation values of
these Hamiltonians 
over symmetric product states give either the purity or the pure state fidelity
of $T$. This allows us to analytically compute these measures for a wide
class of channels, and to identify states that are optimal with respect to
these measures.
\end{abstract}

\maketitle

\section{Introduction}

The study of open quantum systems \cite{Breuer:book} is of interest in
fields as diverse as quantum information science \cite{Nielsen:book},
quantum control \cite{Brumer:book}, and foundations of quantum physics
\cite{Peres:book}. Let $T\in \mathrm{{CP}(\mathcal{H})}$ be a
completely positive 
trace-preserving quantum map i.e., a \emph{channel} over the
finite-dimensional quantum state-space $\mathcal{H}$. The channel $T$ has a
(non-unique) Kraus operator sum representation \cite{Kraus:83} 
\begin{equation}
T(X)=\sum_{i}A_{i}\,X\,A_{i}^{\dagger },\quad \lbrack X\in \mathrm{End}(H)]
\label{kraus}
\end{equation}
where the Kraus operators $A_{i}$ satisfy the constraint
$\sum_{i}A_{i}^{\dagger }A_{i}=\openone$, which guarantees
preservation of 
the trace of a state (density operator) $X=\rho $. A fundamental property of
a state is its purity $p[\rho ]=\mathrm{Tr}(\rho ^{2})$. States are called
pure iff $p=1$ and mixed if $p<1$. In the paradigmatic scenario of open
quantum systems, a state starts out as pure, $\rho =|\psi \rangle \langle
\psi |$, and is then mapped, e.g., via the interaction with an environment,
to a mixed state by the action of a channel $T$:\ $p[T(\rho )]=\mathrm{Tr}
[T^{2}(|\psi \rangle \langle \psi |)]<1$. In this case we say that the state 
$|\psi \rangle $ has been \emph{decohered} by the channel. A typical goal
of, e.g., quantum information processing, is to maximize the purity of a
state that is transmitted via some channel $T$. To this end a variety of
decoherence-reduction techniques have been developed, such as quantum error
correcting codes (QECCs) \cite{Shor:95,Knill:97b,Gottesman:97,Steane:99} and decoherence-free
subspaces (DFSs) \cite{Zanardi:97a,LidarWhaley:03}. In this work we are
interested in the intrinsic purity of quantum channels:

\noindent {\bf{Definition 1}}
The \emph{purity} of the channel $T$ over the subspace $\mathcal{C}\subset 
\mathcal{H}$ is
\begin{equation}
P(T,\mathcal{C}):=\min_{|\psi \rangle \in \mathcal{C}}\mathrm{Tr}
[T^{2}(|\psi \rangle \langle \psi |)].  \label{purityC}
\end{equation}

Minimization is required since we must consider the worst-case scenario. We
invoke subspaces in our definition since we know from the theory of QECCs
and DFSs that it is possible to encode quantum information in a manner that
maximizes purity by restricting to a subspace. In particular:

\noindent {\bf{Definition 2}}
If $P(T,\mathcal{C})=1$ we say that that $\mathcal{C}$ is a decoherence-free
subspace with respect to $T$, in short a $T$-DFS.

In many cases it will not be possible to find a $T$-DFS. A central question
we shall be concerned with here is the characterization of those states that
optimally approximate a $T$-DFS, i.e., those states for which
$P(T,\mathcal{C})$ is as large as possible. Thus: 

\noindent {\bf{Definition 3}}
The optimal purity of $T$ is 
\begin{equation}
P(T):=\max_{\mathcal{C}\subset \mathcal{H}}P(T,\mathcal{C})  \label{purity}
\end{equation}

Note that $P(T)=1\Leftrightarrow $ the set of $T$-DFSs is non-empty.
However, this situation is rather rare and generally requires that there be
a \emph{symmetry} in the system-environment interaction. Associated with
this symmetry is a conserved quantity: quantum coherence. This in turn leads
to the preservation of quantum information. Here we wish to depart from the
notion of a strict symmetry and explicitly consider the situation where one
can only expect optimal, as opposed to ideal purity. However, the
optimization problem defined by $P(T)$ is a hard one, since it involves a
search over all possible subspaces $\mathcal{C}\subset \mathcal{H}$; the
number of such subspaces grows combinatorially in the dimension of
$\mathcal{H}$, which itself may be exponential in the number of
particles, in a typical 
quantum information processing application. Moreover, even if one restricts
the problem to the computation of $P(T,\mathcal{C})$ (for a given, fixed
subspace), one is still faced with a complicated-looking functional.

In this work we focus on the the computation of $P(T,\mathcal{C})$ and
\emph{we associate a Hamiltonian to each channel}. This ``channel
Hamiltonian'' is a mathematical trick, 
rather than a physical Hamiltonian. But, as we shall show,
this has the advantage that it allows us to cast the purity problem
into the familiar framework of computing eigenvalues of Hermitian operators.
In addition, we show that our channel Hamiltonian leads to an elegant
physical (re-) interpretation of the channel purity in terms of the
expectation value of the \textsc{SWAP} operator.

Our work is also related to questions about channel capacity; indeed
recently it has been shown that multiplicativity of generalized maximal
purities implies additivity of the minimal output entropy of the quantum
channel. The latter in turn is equivalent to the additivity of the Holevo
channel capacity \cite{Shor:03}.

We introduce the first channel Hamiltonian in Section~\ref{Hamiltonian}. We then
derive a number of properties and bounds on the purity based on this
formalism in Section~\ref{Properties}. We then devote Section~\ref{Examples}
to a number of examples designed to illustrate our formalism, and derive
some interesting properties for a class of channels. In Section~\ref{Dual}
we derive an alternative interpretation of the expression for the channel
purity, in terms of a dual map. It turns out that the same methods we
introduce for the channel purity also apply to the pure state fidelity of the
channel. In particular, we can introduce a second channel Hamiltonian
to this end. This is addressed in Section~\ref{Fidelity}. We conclude in Section~\ref{Conclusions}.

\section{A Hamiltonian Operator for Quantum Channels}

\label{Hamiltonian}

Associated to the channel $T$ we define an operator over
$\mathcal{H}^{\otimes \,2}$:

\noindent {\bf{Definition 4}} The channel purity-Hamiltonian is:
\begin{equation}
\Omega (T):=\sum_{ij}\Omega _{ij}^{\dagger }\otimes \Omega _{ij},\quad
\Omega _{ij}=A_{i}^{\dagger }A_{j}.  \label{Omega}
\end{equation}

(We shall refer to $\Omega (T)$ simply as the ``channel Hamiltonian''
until our discussion of the pure state fidelity in Section~\ref{Fidelity}.)
It follows immediately from $\Omega _{ij}^{\dagger }=\Omega _{ji}$
that $\Omega (T)$ is a symmetric, Hermitian operator. \emph{Thus
}$\Omega (T)$ \emph{has the status of a Hamiltonian over} $\mathcal{H}^{\otimes \,2}$.
Moreover, $\Omega (T)$ is independent of the particular Kraus operator sum
representation chosen for $T$: all possible operator
sum-representations of $T$ are obtained by considering new Kraus
operators of the form $A_{i}^{\prime }=\sum_{j}U_{ij}A_{j}$ where the $U_{ij}$'s are the entries of
unitary matrix. By inserting this expression into the definition (\ref{Omega}) one can explicitly check that $\Omega (T)$ is invariant.

We now come to
our key result: a representation of the purity of quantum channels as the
expectation value of the channel Hamiltonian:

\noindent {\bf{Proposition 0}}
For every quantum channel $T$ and subspace $\mathcal{C}$ one has the identity 
\begin{equation}
P(T,\mathcal{C})=\min_{|\psi \rangle \in \mathcal{C}}\langle \psi ^{\otimes
\,2}|\Omega (T)|\psi ^{\otimes \,2}\rangle .  \label{purityOm}
\end{equation}

\begin{proof}
One has 
\begin{align}
& \mathrm{Tr}T^{2}(|\psi \rangle \langle \psi |)=\sum_{ij}\langle \psi
|A_{j}^{\dagger }A_{i}|\psi \rangle \langle \psi |A_{i}^{\dagger }A_{j}|\psi
\rangle  \notag \\
& =\sum_{ij}|\langle \psi |A_{j}^{\dagger }A_{i}|\psi \rangle |^{2}=\mathrm{Tr}[|\psi \rangle \langle \psi |^{\otimes \,2}A_{j}^{\dagger }A_{i}\otimes
A_{i}^{\dagger }A_{j}]  \notag \\
& =\mathrm{Tr}[|\psi \rangle \langle \psi |^{\otimes \,2}\Omega (T)].
\end{align}
Equation (\ref{purityOm}) now follows by taking the minimum over $|\psi \rangle
\in \mathcal{C}$.
\end{proof}

Note that $\Omega (T)$ is a formal Hamiltonian over the \textquotedblleft
double\textquotedblright\ Hilbert space $\mathcal{H}^{\otimes \,2}$, and is therefore
unrelated to the physical Hamiltonian for the original problem. However, as
we show below, there does exist an attractive physical interpretation
of Eq.~\ref{purityOm}, in terms of the expected value of the
\textsc{SWAP} operator.

\section{Bounds and Other Channel Properties}

\label{Properties}

We now derive upper and lower bounds on the purity and then give a
characterization of $T$-DFSs.

\noindent {\bf{Proposition 1}}
Let $\omega _{0}^{+}(A)$ denote the minimum eigenvalue of the symmetric
operator $A$ in the symmetric subspace of $\mathcal{H}^{\otimes \,2}$, and
let $\Pi ^{+}(\mathcal{C})$ denote the normalized projector over the
symmetric part of of $\mathcal{C}^{\otimes \,2}$. Then the following bounds
hold:
\begin{equation}
\mathrm{Tr}[\Pi ^{+}(C)\Omega (T)]\geq P(T,C)\geq \omega
_{0}^{+}[\Omega (T)] .
\label{bounds}
\end{equation}

\begin{proof}
Note that since $\Omega (T)$ is a symmetric operator, the symmetric subspace
of $\mathcal{H}^{\otimes \,2}$ is $\Omega (T)$-invariant. Therefore the
minimum expectation value of $\Omega (T)$ in this subspace coincides with
the minimum eigenvalue $\omega _{0}^{+}$. The lower bound in Eq.~(\ref{bounds}) is simply due to the fact that minimization over the symmetric
subspace of $\mathcal{H}^{\otimes \,2}$ includes the minimization over
the product states $|\psi \rangle ^{\otimes \,2}\in
\mathcal{C}^{\otimes \,2}$. The upper bound
in Eq. (\ref{bounds}) derives from the identity $\int_{\mathcal{C}}|\psi
\rangle \langle \psi |^{\otimes \,2}=\Pi ^{+}(\mathcal{C})$ (integration
over the uniform distribution over $\mathcal{C}$ \cite{Zanardi:00}) and from the
obvious fact that the average value of a function is no smaller than its
minimum value.
\end{proof}

\noindent {\bf{Lemma 1}}
Let $T$ be unital [$T(\openone)=\openone$]. Then $\forall |\psi \rangle \in 
\mathcal{H}$ we have:
\begin{equation}
\Vert \Omega (T)|\psi \rangle ^{\otimes \,2}\Vert \leq 1.
\end{equation}

\begin{proof}
Let $p_{ij}:= || A_{j}^{\dagger }A_{i}|\psi \rangle || $.
One has the following normalization condition 
\begin{equation*}
\sum_{ij}p_{ij}^{2}=\sum_{ij}\langle \psi |A_{i}^{\dagger
}A_{j}A_{j}^{\dagger }A_{i}|\psi \rangle =\sum_{i}\langle \psi
|A_{i}^{\dagger }A_{i}|\psi \rangle =1,
\end{equation*}
where in the first (second) equality we used the unitality (CP-map)
condition $\sum_{j}A_{j}A_{j}^{\dagger }=\openone$ ($\sum_{i}A_{i}^{\dagger
}A_{i}=\openone$). Now 
\begin{eqnarray}
\Vert \Omega (T)|\psi \rangle ^{\otimes \,2}\Vert &=&\Vert
\sum_{ij}A_{j}^{\dagger }A_{i}|\psi \rangle \otimes A_{i}^{\dagger
}A_{j}|\psi \rangle \Vert  \notag \\
&\leq &\sum_{ij}\Vert A_{j}^{\dagger }A_{i}|\psi \rangle \Vert \,\Vert
A_{i}^{\dagger }A_{j}|\psi \rangle \Vert =\sum_{ij}p_{ij}p_{ji}  \notag \\
&\leq &\sum_{ij}p_{ij}^{2}=1,
\end{eqnarray}
where in the last line we used the Cauchy-Schwartz inequality for the
Hilbert-Schmidt product of matrices, 
\begin{eqnarray*}
\sum_{ij}p_{ij}p_{ji}&=&\mathrm{Tr}P^{2}=\langle P,P^{\dagger }\rangle \leq
\Vert P\Vert \,\Vert P^{\dagger }\Vert \\
&=& \Vert P\Vert
^{2}=\sum_{ij}p_{ij}^{2}.
\end{eqnarray*}
\end{proof}

We now proceed to characterize $T$-DFSs. To this end we introduce a special
subspace:

\noindent {\bf{Definition 5}}
The subspace ${\mathcal H}^{\Omega }$ of $\Omega $-invariant states (${\mathcal H}^{\Omega
}\subset {\mathcal H}^{\otimes \,2}$) is the eigenspace of $\Omega $ with eigenvalue
one.

\noindent {\bf{Proposition 2}}

\begin{itemize}
\item[i)] If $\forall |\psi \rangle \in \mathcal{C}$ and $\forall i$ it
holds that: $A_{i}|\psi \rangle =\alpha _{i}U|\psi \rangle ,\,A_{i}^{\dagger
}|\psi \rangle =\alpha _{i}^{\ast }U^{\dag }|\psi \rangle $, where $U$ is
unitary, then $|\psi \rangle ^{\otimes \,2}\in H^{\Omega }$.

\item[ii)] Let $T$ be unital. Then $\mathcal{C}$ is a $T$-DFS
  $\Leftrightarrow \mathcal{C}^{\otimes \,2}\subset
  \mathcal{H}^{\Omega }$.

\item[iii)] $T$-DFS $\Leftrightarrow $ the first inequality in Eq.~(\ref{bounds}) is an equality.
\end{itemize}

\begin{proof}
i) Notice first that from the CP-map condition, $\sum_{i}A_{i}^{\dagger
}A_{i}=\openone$, it follows that $\sum_{i}|\alpha _{i}|^{2}=1.$ Now for $
|\psi \rangle \in \mathcal{C}$ one has that $\langle \psi ^{\otimes
\,2}|\Omega (T)|\psi ^{\otimes \,2}\rangle =\sum_{ij}|\alpha _{i}\alpha
_{j}|^{2}=(\sum_{i}|\alpha _{i}|^{2})^{2}=1.$

ii) ($\Rightarrow $) If $\mathcal{C}$ is a $T$-DFS then $\min_{|\psi \rangle
\in \mathcal{C}}\langle \psi ^{\otimes \,2}|\Omega (T)|\psi ^{\otimes
\,2}\rangle =1$. But from the Cauchy-Schwartz inequality and Lemma~1 above
one has that $\langle \psi ^{\otimes \,2}|\Omega (T)|\psi ^{\otimes
\,2}\rangle \leq 1$ $(\forall |\psi \rangle )$ and the equality holds
iff $\Omega (T)|\psi ^{\otimes \,2}\rangle =|\psi ^{\otimes
  \,2}\rangle ~(\forall 
|\psi \rangle \in \mathcal{C})$. Now, if $|\Psi \rangle $ is in the
symmetric part of $\mathcal{C}^{\otimes \,2}$, one has that 
$|\Psi \rangle=\Pi_+({\cal C})|\Psi \rangle=
\alpha({\cal C})   \int_{\mathcal{C}}|\psi ^{\otimes \,2}\rangle \langle \psi ^{\otimes
\,2}|\Psi \rangle$ [where $\alpha({\cal C}):= {\rm dim}{\cal C}({\rm dim}{\cal C}+1)$]; therefore 
\begin{eqnarray}
\Omega (T)|\Psi \rangle &=&\alpha({\cal C}) 
\int_{\mathcal{C}}\Omega (T)|\psi ^{\otimes \,2}\rangle \langle \psi ^{\otimes \,2}|\Psi
\rangle
\nonumber\\ 
& =&\alpha({\cal C}) \int_{\mathcal{C}}|\psi ^{\otimes \,2}\rangle \langle \psi
^{\otimes \,2}|\Psi \rangle 
.\end{eqnarray}
It follows that $|\Psi \rangle \in H^{\Omega }$.

($\Leftarrow $) If $\forall |\psi \rangle \in \mathcal{C}$ it holds that $|\psi \rangle ^{\otimes \,2}\in \mathcal{H}_{1}^{\Omega }$, then $\mathcal{C}$ is
clearly a $T$-DFS, i.e., $P(T,\mathcal{C})=1$. A fortiori this holds if all
the elements of the symmetric part of $\mathcal{C}^{\otimes \,2}$ are in $\mathcal{H}^{\Omega }$.

iii) We have just seen that $\forall |\Psi \rangle =|\psi \rangle ^{\otimes
\,2}$ such that $|\psi \rangle \in \mathcal{C}$ one has $\langle \Psi
|\Omega (T)|\Psi \rangle =1$. By integrating over $|\psi \rangle $ one
obtains that the average [leftmost part of Eq. (\ref{bounds})] coincides
with the minimum [middle term in (\ref{bounds})].
\end{proof}

\section{Examples}

\label{Examples}

We now present a variety of examples to illustrate our formalism, to actually compute the purity of a number of interesting
channels, and to find the corresponding optimally pure states.

\noindent {\bf{Example 1}} 
 \label{ex1}{\em Single qubit anisotropic depolarizing channel}.
 
Let $T$ be the
one-qubit channel given by $\rho \rightarrow \sum_{i=0}^{3}p_{i}\sigma
^{i}\rho \sigma ^{i}$ where the $\sigma ^{i}$'s are the Pauli matrices
($\sigma ^{0}=\openone$) and the $p_{i}$'s a probability
distribution. One 
finds $\Omega _{01}=\Omega _{10}=\sqrt{p_{0}p_{i}}\sigma ^{0}\sigma
^{i}= \sqrt{p_{0}p_{i}}\sigma ^{i}$; $\Omega _{ij}=-\Omega _{ji}=i\sqrt{p_{j}p_{i}}%
\epsilon _{ijk}\sigma ^{k}$ $\,(i=1,2,3)$; $\Omega
_{ii}=p_{i}\openone$, $(i=0,\ldots ,3)$. It then follows that 
\begin{equation}
\Omega (T)=\sum_{i=0}^3 \alpha _{i}\sigma ^{i}\otimes \sigma ^{i}  \label{OmOne}
\end{equation}
where $\alpha _{0}=\sum_{i=0}^{3}p_{i}^{2}$,$\,\alpha
_{k}=2(p_{0}p_{k}+p_{i}p_{j})$ $(i\neq j\neq k,k=1,2,3)$. Note that $
\sum_{i=0}^{3}\alpha _{i}=(\sum_{i=0}^{3}p_{i})^{2}=1$, that $\alpha _{0}\in
\lbrack 1/4,1]$,$\,$and $\alpha _{k}\in \lbrack 0,1/2]$,$\,(k=1,2,3)$. The
eigenstates of $\Omega (T)$ are the Bell states $|\psi ^{-}\rangle
=(|01\rangle -|10\rangle )/\sqrt{2}$ (singlet) and $\{|\phi ^{-}\rangle
=(|00\rangle -|11\rangle )/\sqrt{2},|\psi ^{+}\rangle =(|01\rangle
+|10\rangle )/\sqrt{2},|\phi ^{+}\rangle =(|00\rangle +|11\rangle
)/\sqrt{2}\}$ (triplet). Their respective eigenvalues are $2\alpha
_{0}-1$ and $\{1-2\alpha _{1},1-2\alpha _{3},1-2\alpha
_{2}\}$. Further note that Spec$\Omega (T)\subset \lbrack -1/2,\,1]$ and that $\Omega (T)$ in the triplet
sector is a positive operator. The triplet sector is symmetric, while the
singlet is anti-symmetric. From Eq.~(\ref{bounds}) we thus know that the
minimal eigenvalue in the triplet sector provides a lower bound on the
purity:
\begin{equation}
P(T)\geq 1-2\max_{i=1,2,3}\alpha _{i}\geq 0
\end{equation}
(of course in this single-qubit example there are no non-trivial
subspaces: $\mathcal{C}=\mathcal{H}$). In this general case we cannot
directly determine 
the actual purity or find the corresponding maximally robust state(s), since
the Bell triplet-states are not product states. To find the optimal purity
states in such a case one has to resort to other optimization techniques.
However, in certain special cases the eigenstates of $\Omega (T)$ will be
product states, whence our method directly yields the optimally robust
states. For instance, consider the case $p_{0}=p_{1}=1/2$ and
$p_{2}=p_{3}=0$; one finds $\Omega (T)=(\openone+\sigma _{x}\otimes
\sigma _{x})/2$. In 
this case $|\psi ^{-}\rangle ,|\phi ^{-}\rangle $ are degenerate, as are $
|\psi ^{+}\rangle ,|\phi ^{+}\rangle $. We then find, respectively, the
symmetric product eigenstates $[(|0\rangle -|1\rangle )/\sqrt{2}]^{\otimes 2}
$ and $[(|0\rangle +|1\rangle )/\sqrt{2}]^{\otimes 2}$, both with eigenvalue 
$1$. The states $|\pm \rangle :=(|0\rangle \pm |1\rangle )/\sqrt{2}$ are
thus both $T$-DFSs. This is intuitively clear, as the channel in this case
is simply $T(\rho )=p_{o}\rho +p_{1}\sigma ^{x}\rho \sigma ^{x}$, and the
states $|\pm \rangle $ lie on the Bloch sphere $x$-axis, which is invariant.

As another example, consider the fully depolarizing channel with $%
p_{i}=(1-p_{0})/3$,$\,(i=1,2,3)$. Then the following (antiferromagnetic
Heisenberg exchange) Hamiltonian is obtained: $\Omega (T)=\alpha _{0}\openone
+\alpha \sum_{i=1}^{3}\sigma ^{i}\otimes \sigma ^{i}$ where $\alpha
_{0}=p_{0}^{2}+(1-p_{0})^{2}/3$,$\,\alpha
=(2/3)[p_{0}(1-p_{0})+(1-p_{0})^{2}/3]\geq 0$. We can rewrite this as $
\Omega (T)=\alpha _{0}\openone+\alpha (2S-\openone)$, where the \textsc{SWAP}
operator $S$ is defined by its action on basis states: $S|i\rangle \otimes
|j\rangle =|j\rangle \otimes |i\rangle $. In this case clearly every
symmetric product state is an eigenstate of $\Omega (T)$, with
eigenvalue $\alpha _{0}+\alpha $, which equals the channel
purity. Thus all single-qubit
states are equally (and optimally) robust. Again, this is intuitively
clear: the fully depolarizing channel shrinks the Bloch sphere
isotropically.

\noindent {\bf{Example 2}} 
 \label{ex2}
  {\em Correlated two-qubit anistropic depolarizing channel}.

Consider the correlated map
\begin{equation}
T(\rho )=\sum p_{\alpha }(\sigma _{\alpha }\otimes \sigma _{\alpha })\rho
(\sigma _{\alpha }\otimes \sigma _{\alpha }).
\end{equation}%
Then $\Omega _{\alpha \beta }=\sqrt{p_{\alpha }p_{\beta }}\sigma _{\alpha
}\sigma _{\beta }\otimes \sigma _{\alpha }\sigma _{\beta }$ and
\begin{equation}
\Omega =\sum_{\alpha =0,x,y,z}p_{\alpha }^{2}(I\otimes I)^{\otimes
2}+\sum_{\alpha \neq \beta \neq \gamma }p_{\alpha }p_{\beta }(\sigma
_{\gamma }\otimes \sigma _{\gamma })^{\otimes 2}.
\end{equation}
This example can be solved directly by observing that each of the Bell
states has eigenvalue $+1$ or $-1$ under the action of $\sigma_\gamma
\otimes \sigma_\gamma$, whence the purity is one. Thus the Bell states
are $T$-DFSs. 
Notice that this result appears to be related to the communication problem
for channels with correlated noise studied in Ref.~\cite{mapa}.

One can
also find the {\em minimal} purity states by differentiating $\langle \Omega
\rangle := \,^{\otimes 2}\langle \psi |\Omega(T)|\psi\rangle^{\otimes
  2}$ as a function of expansion parameters of $\ket{\psi}$ over the
Bell states. This yields $\langle \Omega \rangle _{\min
}=\sum_{\alpha =0,x,y,z}p_{\alpha }^{2}$, and the corresponding minimally
robust set of states are superpositions of pairs of Bell states with
arbitrary phases:
\begin{align*}
|\psi _{1}\rangle & =\frac{e^{i\alpha _{1}}}{2}(|00\rangle +|11\rangle )+
\frac{e^{i\beta _{1}}}{2}(|01\rangle -|10\rangle ) \\
|\psi _{2}\rangle & =\frac{e^{i\alpha _{2}}}{2}(|00\rangle -|11\rangle )+
\frac{e^{i\beta _{2}}}{2}(|01\rangle +|10\rangle ).
\end{align*}
This channel thus has the interesting property that the maximally entangled
Bell states are more robust than any separable (pure) state.

\noindent {\bf{Example 3}} 
 {\em Amplitude damping}.
 
Let $T(\rho )=|0\rangle \langle 0|$, $\forall \rho \in S({\CC}^{d})$. A
  set of Kraus operators is given by $A_{i}=|0\rangle \langle i|,\,(i=1,\ldots ,d)$. Note that the channel is non-unital: $\sum_{i}A_{i}A_{i}^{\dagger }=d\,|0\rangle \langle 0|>\openone$. One has $
A_{j}^{\dagger }A_{i}=|j\rangle \langle i|$, so that $\Omega
(T)=\sum_{ij}|j\rangle \langle i|\otimes |i\rangle \langle
j|=\sum_{ij}|ji\rangle \langle ij|=S$ (the {\sc SWAP} operator). Here all states are mapped onto a
pure one and $\Omega (T)$ is identically $\openone$ in the symmetric
subspace. A slight generalization is given by $T(\rho )=(1-p)\rho
+p|0\rangle \langle 0|$,$\,(p\in \lbrack 0,1])$. In this case one finds 
\begin{eqnarray}
\Omega (T) &=&(1-p)^{2}\openone+p^{2}S  \notag \\
&+&p(1-p)(|0\rangle \langle 0|\otimes \openone+\openone\otimes |0\rangle
\langle 0|)S
\end{eqnarray}
Note that $\Vert \Omega (T)\Vert \leq (1-p)^{2}+p^{2}+2p(1-p)=1$. The only $T
$-DFS is ${\CC}|0\rangle$.

\noindent {\bf{Example 4}} 
{\em  Projective measurements}.

Let $T(\rho )=\sum_{i}\Pi _{i}\rho \Pi _{i}$,$%
\;\Pi _{i}\Pi _{j}=\delta _{ij}\Pi _{i}$,$\,\sum_{i}\Pi _{i}=\openone$. Then 
$\Omega _{ij}=\delta _{ij}\Pi _{i}$, from which 
\begin{equation}
\Omega (T)=\sum_{i}\Pi _{i}\otimes \Pi _{i}.  \label{OmPro}
\end{equation}%
If $\mathcal{H}_{i}:={\rm Im}\Pi _{i}$ then $\mathcal{H}_{i}^{\otimes
\,2}\subset \mathcal{H}^{\Omega }$, i.e., from Proposition~2 all
the eigenvectors of the $\Pi _{i}$'s are 1D $T$-DFSs. The maximum
eigenvalue of $\Omega (T)$ is $1$; this follows from $\langle \Psi |\Omega (T)|\Psi
\rangle \leq 1\,$($\forall |\Psi \rangle $). The latter inequality results
from the following argument: $\openone^{\otimes \,2}=(\sum_{i}\Pi
_{i})^{\otimes \,2}=\sum_{ij}\Pi _{i}\otimes \Pi _{j}=\Omega (T)+\sum_{i\neq
j}\Pi _{i}\otimes \Pi _{j}.$ The last term is a non-negative operator (sum
of products of non-negative operators), whence $\openone-\Omega (T)\geq 0$.
Taking the expectation value of the last inequality with respect to $|\Psi
\rangle $ proves the bound above.

In the following we use the operator norm $\left\Vert A\right\Vert _{\infty
}:=\max_{\left\Vert |\psi \rangle |\right\Vert =1}\left\Vert A|\psi \rangle
\right\Vert $. We shall write $\left\Vert A\right\Vert $ for simplicity.

\noindent {\bf{Example 5}}  
{\em Unitary mixture of a group representation}.

This is a rather general and
quite important example, which includes Examples~1,2 above.
Let 
\begin{equation}
T(\rho )=\sum_{g}p_{g}U_{g}\rho U_{g}^{\dagger }
\end{equation}
where $g\mapsto p_{g}$ is a probability distribution over the group
$\mathcal{G}=\{g\}$ and $g\mapsto U_{g}$ a unitary representation of
$\mathcal{G}$. 
One finds $\Omega _{gh}=\sqrt{p_{g}p_{h}}U_{g}^{\dagger }U_{h}=\sqrt{
p_{g}p_{h}}U_{g^{-1}h}$; thus 
\begin{equation}
\Omega (T)=\sum_{k\in \mathcal{G}}q_{k}U_{k}\otimes U_{k^{-1}},
\label{OmGru}
\end{equation}
where $q_{k}:=\sum_{g}p_{g}p_{gk^{-1}}$ is also a probability distribution.
If $|\psi \rangle \in \mathcal{H}$ is a $\mathcal{G}$-singlet, i.e., $U_{g}|\psi
\rangle =|\psi \rangle$ ($\forall g\in G$) then $|\psi \rangle ^{\otimes
\,2}\in \mathcal{H}^{\Omega }$, i.e., all the $\mathcal{G}$-singlets are 1D $T$-DFSs.
Here again, the maximum eigenvalue of $\Omega (T)$ is $1$: Indeed,
it is easy to see that $\Vert \Omega (T)\Vert \leq 1$: $\Vert \Omega
(T)\Vert \leq \sum_{k}q_{k}\Vert U_{k}\otimes U_{k^{-1}}\Vert
=\sum_{k}q_{k}=1$.

As a particular instance of this kind of channel let us consider an $N$-qubit case with the $U_{k}$'s generating an Abelian subgroup of $\mathcal{G}
$ the Pauli group (all tensor products of Pauli matrices on $N$ qubits), as
was the case in Examples~1,2 above. The set of $\mathcal{G}$-singlets
is now given by the the stabilizer of $\mathcal{G}$ [denoted ${\rm Stab}(\mathcal{G})$], i.e., the subspace generated by the $|\psi \rangle $ such
that $U_{k}|\psi \rangle =|\psi \rangle \,$($\forall k$)$.$ Since
$U_{k}=U_{k}^{\dagger }$ one finds immediately that elements of the
form $|\psi \rangle ^{\otimes \,2}$, where $|\psi \rangle \in
{\rm Stab}(\mathcal{G})$,
are eigenvectors of $\Omega (T)$ with maximum eigenvalue $\sum_{k}q_{k}$.
These states also play the role of codewords of stabilizer QECCs \cite{Gottesman:97}. We thus see that, in this example, the stabilizer-QECC
codewords are maximally robust, though no active error correction is assumed.

In fact, the connection to quantum error correction can be made more
general: the formalism developed so far allows us to establish an
intriguing identity for the purity of states belonging to a QECC $\mathcal{C}
$ for the $CP$-map $T$: $\rho \rightarrow \sum_{i}A_{i}\rho A_{i}^{\dagger }$. If $|\psi _{\alpha }\rangle ,|\psi _{\beta }\rangle \in \mathcal{C}$ then
the error-correction condition is:
\begin{equation}
\langle \psi _{\alpha }|A_{i}^{\dagger }A_{j}|\psi _{\beta }\rangle
=c_{ij}\delta _{\alpha \beta },
\end{equation}
where the matrix $c_{ij}$ is Hermitian, non-negative, and has trace one \cite{Knill:97b}. For non-degenerate codes $c_{ij}$ has maximal rank. Let us now
consider states of the form $|\psi _{\alpha }\rangle ^{\otimes \,2}\,$($%
|\psi _{\alpha }\rangle \in \mathcal{C}$). From Eq.~(\ref{Omega}) and the
error correction condition one has that 
\begin{eqnarray}
P(T,\mathcal{C}) &=&\langle \psi _{\alpha }^{\otimes \,2}|\Omega (T)|\psi
_{\alpha }^{\otimes \,2}\rangle =\sum_{ij}|\langle \psi _{\alpha
}|A_{i}^{\dagger }A_{j}|\psi _{\alpha }\rangle |^{2} \nonumber \\
&=&\sum_{ij}|c_{ij}|^{2}=\mathrm{Tr}(c^{2}).
\end{eqnarray}
Viewing $c$ as a state (density operator), we have thus found that \emph{the
purity of the channel acting on the codewords of }$\mathcal{C}$\emph{\ is
just the purity of the \textquotedblleft state\textquotedblright\
}$c$\emph{\ associated to the code itself}. For example, for DFSs $c$
is simply a rank 
one matrix with unit trace \cite{Lidar:PRL99}, whence $\mathrm{Tr}c^{2}=1$
and the maximum eigenvalue condition is readily recovered. As a more
interesting example, consider a CP-map $T$ with $A_{i}=\sqrt{p_{i}}U_{i}$
with unitary $U_{i}$'s (e.g., chosen from the Pauli group, as in stabilizer
QEC). Recall that $c=\lambda ^{\dag }\lambda $, where the matrix $\lambda $
is defined by the error-recovery relation $R_{r}A_{i}=\lambda _{ri}\openone$
(restricted to $\mathcal{C}$), for each recovery operator $R_{r}$ \cite{Knill:97b}.
Then $R_{r}=\lambda _{ri}U_{i}^{-1}/\sqrt{p_{i}}$, and from the CP-condition 
$\sum_{r}R_{r}^{\dag }R_{r}=\openone$ we find $\sum_{r}|\lambda
_{ri}|^{2}=p_{i}$. Assume for simplicity that there is a unique recovery
operator per error, i.e., $\lambda _{ri}=\lambda _{i}\delta _{ri}$, $\lambda
_{i}$ $\neq 0$ $\forall i$ (this is an example of a non-degenerate code).
Then $|\lambda _{i}|^{2}=p_{i}$ and $c=\mathrm{diag}(p_{i})$; it follows
that the purity over such a QECC associated to $T$ is simply given by $%
\sum_{i}p_{i}^{2}$.

\section{The Dual Representation}
\label{Dual}

We now develop an alternative representation of the channel-Hamiltonian,
which is useful for the derivation of several additional results, and sheds
new light on the physical interpretation of the channel purity.

\noindent {\bf{Definition 6}}
The dual $T_{\ast }$ of a CP-map $T$ [see Eq (\ref{kraus})] is: $T_{\ast
}(X)=\sum_{i}A_{i}^{\dagger }XA_{i}.$

\noindent {\bf{Proposition 3}}
Let $S$ be the \textsc{SWAP}\ operator (defined above). Then: 
\begin{equation}
\Omega (T)=T_{\ast }^{\otimes \,2}(S)S.  \label{light}
\end{equation}

We give two different proofs.

\begin{proof}
a) 
\begin{eqnarray}
T_{\ast }^{\otimes 2}(S)S &=&\sum_{ij}(A_{i}^{\dagger }\otimes
A_{j}^{\dagger })S(A_{i}\otimes A_{j})S  \notag \\
&=&\sum_{ij}(A_{i}^{\dagger }\otimes A_{j}^{\dagger })(A_{j}\otimes A_{i}) 
\notag \\
&=&\sum_{ij}A_{i}^{\dagger }A_{j}\otimes A_{j}^{\dagger }A_{i}=\Omega (T)
\end{eqnarray}

b) By writing the \textsc{SWAP}\ operator explicitly as $S=\sum_{lm}|m%
\rangle \langle l|\otimes |l\rangle \langle m|$ and applying $T_{\ast
}^{\otimes \,2}$ one obtains: $\sum_{lm,ij}A_{i}^{\dagger }|m\rangle \langle
l|A_{i}\otimes A_{j}^{\dagger }|l\rangle \langle m|A_{j}.$ Then the proof
follows by explicity comparing the matrix elements of the latter operator
times $S$ with the ones of $\Omega (T)$.
\end{proof}

We remark that one is led to consider the operator $T_{\ast }^{\otimes
\,2}(S)$ by the following argument: 
\begin{eqnarray*}
\mathrm{Tr}[T^{2}(\rho )] &=&\mathrm{Tr}[S\{T(\rho )\otimes T(\rho )\}] \\
&=&\mathrm{Tr}[ST^{\otimes \,2}(\rho \otimes \rho )]=\mathrm{Tr}[T_{\ast
}^{\otimes \,2}(S)\rho \otimes \rho ],
\end{eqnarray*}%
where in the first step we used the identity
\begin{equation}
\mathrm{Tr}[AB] =\mathrm{Tr}[S A\otimes B] 
\end{equation}
valid for general operators $A,B$ \cite{proof}, and in the last step we ``dualized'' the
map. Then for pure inputs $\rho =|\psi \rangle \langle \psi |$ one has $%
\langle \psi ^{\otimes \,2}|\Omega (T)|\psi ^{\otimes \,2}\rangle =\langle
\psi ^{\otimes \,2}|T_{\ast }^{\otimes \,2}(S)|\psi ^{\otimes \,2}\rangle $.
This dualization is quite useful since it moves the burden of calculation of
the channel action away from the entire \emph{set} of states $\rho $ to the 
\emph{single} observable $S$.

\noindent {\bf{Corollary 1}}
Upon restriction to the to the symmetric subspace of $\mathcal{H}^{\otimes
\,2}$ one can write $\Omega (T)=T_{\ast }^{\otimes \,2}(S)$.

\begin{proof}
Immediate.
\end{proof}

The following corollary contains a general derivation, based on the dual representation of $\Omega(T)$,
of a fact that was already proved for specific examples in Section~\ref{Examples}.

\noindent {\bf{Corollary 2}}
$\Vert \Omega (T)\Vert \leq 1.$

\begin{proof}
One has $\Vert \Omega (T)\Vert =\Vert T_{\ast }^{\otimes \,2}(S)S\Vert \leq
\Vert T_{\ast }^{\otimes \,2}(S)\Vert \,\Vert S\Vert \leq \Vert T_{\ast
}^{\otimes \,2}(S)\Vert $. Since $T_{\ast }^{\otimes \,2}$ is the dual of a
CP map elements smaller (greater) than the identity (minus the identity) are
mapped onto elements smaller than the identity. Since $-\openone\leq S\leq 
\openone$ one has $-\openone\leq T_{\ast }^{\otimes \,2}(S)\leq \openone$.
This relation implies in particular that the maximum eigenvalue of the
Hermitian operator $T_{\ast }^{\otimes \,2}(S)$ is smaller than one. Since
this maximum eigenvalue coincides with the $\left\Vert {\cdot}\right\Vert
_{\infty }$ norm of $T_{\ast }^{\otimes \,2}(S)$ the inequality is proved.
\end{proof}

We now present a result that allows one to directly compute the {\em average} purity of a
quantum channel.

\noindent {\bf{Proposition 4}}
The Haar average purity of the $CP$-map $T$ is given by 
\begin{equation}
\overline{\mathrm{Tr}[T^{\otimes 2}(|\psi \rangle \langle \psi |)]}^{\psi }=
\frac{1}{d(d+1)}\mathrm{Tr}[ST^{\otimes \,2}(\openone)+\Omega (T)]
\label{average}
\end{equation}

\begin{proof}
Using the fact that $\int d\psi |\psi \rangle \langle \psi |^{\otimes \,2}$
is the normalized projector over the symmetric subspace of $\mathcal{H}%
^{\otimes \,2}$, i.e., $(\openone+S)/d(d+1)$ \cite{Zanardi:00}, one has 
\begin{eqnarray}
&&\int d\psi \mathrm{Tr}\left[ T_{\ast }^{\otimes \,2}(S)|\psi \rangle
\langle \psi |^{\otimes \,2}\right] =\frac{\mathrm{Tr}\left[
T_{\ast }^{\otimes \,2}(S)(\openone+S)\right]}{d(d+1)}  \notag \\
&=&\frac{1}{d(d+1)}\mathrm{Tr}\left[ ST^{\otimes \,2}(\openone)+T_{\ast
}^{\otimes \,2}(S)S\right]
\end{eqnarray}%
\end{proof}

In other words, {\em the Haar-average purity of a channel} $T$ {\em is given by the expectation value
of} $\Omega(T)$ {\em over the maximally mixed state} $\Pi_+({\cal H})=
1/d(d+1) (\openone +S)$ {\em over the symmetric subspace of} ${\cal
  H}^{\otimes\,2}$.
 
\noindent {\bf{Corollary 3}}
Using Eq.~(\ref{average}) one can get the Haar-averaged purities of the
channels considered above:

(i) One qubit depolarizing channel: $(1+2\alpha _{0})/3$.

(ii) Amplitude damping channel: $(1-p)^{2}+p^{2}+2p(1-p)/d$.

(iii) Projective measurements: $[d+\sum_{i}(\mathrm{Tr}\Pi
  _{i})^{2}]/d(d+1)$.

(iv) Unitary mixing: $[d+\sum_{g\in \mathcal{G}}q_{g}|\mathrm{Tr}
U_{g}|^{2}]/d(d+1)$

>From (iii) and (iv) it follows that:

(a) One-dimensional projective measurements achieve the minimal
average purity, of $2/(d+1)$.

(b) For unitary mixing and assuming a Haar-uniform distribution (all $q_{g}$
equal, i.e., the fully depolarizing channel), minimal purity is obtained for 
$U_{g}$'s in a $\mathcal{G}$-irrep. Indeed, one has in general that $(1/|
\mathcal{G}|)\sum_{g\in \mathcal{G}}|\mathrm{Tr}U_{g}|^{2}=\sum_{J}n_{J}^{2}$
where $n_{J}$ is the multiplicity of the $J$-th $\mathcal{G}$ irrep
\cite{Cornwell:84I+II}. The minimum is clearly achived when just one
irrep appears, i.e., the irreducible case.

Before concluding this section we would like to point out that the formula
$\langle\Psi|\Omega(T)|\Psi\rangle={\rm Tr}\left[ ST^{\otimes\,2}(|\Psi\rangle\langle\Psi|)\right]$
allows us to give an operational  meaning to the operator $\Omega$, and in the particular case
in which $|\Psi\rangle=|\psi\rangle^{\otimes\,2}$, to the purity of $T(|\psi\rangle\langle\psi|)$.
Indeed, this expectation value of $\Omega(T)$ is nothing but
the expectation  value of the observable $S$ in the state $T^{\otimes\,2}(|\Psi\rangle\langle\Psi|)$.
The latter state can in turn be viewed as the result of an  action of the channel
on a pair of, possibly entangled,  input states from ${\cal H}$. 

\section{Pure State Fidelity of a Channel}
\label{Fidelity}

We now show how many of the techniques introduced above for the channel
purity carry over to the (simpler) problem of calculating the

\noindent {\bf{Definition 7}} 
 Pure state fidelity:
  \begin{equation}
  F(T,|\psi\rangle):=\langle\psi|T(|\psi\rangle\langle\psi|)|\psi\rangle .
  \end{equation}

\noindent {\bf{Proposition 5}}

(i) 
\begin{equation}
F(T,|\psi\rangle)=\langle
\psi^{\otimes\,2}|\Omega_1(T)|\psi^{\otimes\,2}\rangle
\end{equation}
where $\Omega_1(T)=(\openone\otimes T_*)(S)S$.

(ii)
\begin{equation}
 \Omega_1(T)=\sum_{i}A_i\otimes A_i^\dagger .
\end{equation}

(iii) 
\begin{equation}
\overline{F(T,|\psi\rangle)}^{\psi}
=\frac{1}{d(d+1)}{\rm Tr}\left[ \Omega_1(T) + S(\openone\otimes T(\openone))
  \right ] .
\label{av_fid}
\end{equation}
In particular, for a {\em unital map} one has average pure state fidelity given by
$[ d+\sum_i|{\rm Tr}A_i|^2]/d(d+1)$.

 \begin{proof}

(i) 
\begin{eqnarray}
F(T,|\psi\rangle)&=&\langle\psi|T(|\psi\rangle\langle\psi|)|\psi\rangle={\rm Tr}\left[
|\psi\rangle\langle\psi| T(|\psi\rangle\langle\psi|)\right]
\nonumber\\
 &=&{\rm Tr}\left[S |\psi\rangle\langle\psi| 
\otimes T(|\psi\rangle\langle\psi|)\right]
\nonumber\\
&=&{\rm Tr}\left[S (\openone\otimes
  T)|\psi\rangle\langle\psi|^{\otimes\,2}\right]\nonumber\\
&=&{\rm Tr}\left[(\openone\otimes T_*)(S)|\psi\rangle\langle\psi|^{\otimes\,2}\right]\nonumber\\
&=&
\langle\psi^{\otimes\,2}|(\openone\otimes T_*)(S) S|\psi^{\otimes\,2}\rangle.
\nonumber
\end{eqnarray}

(ii) 
\begin{eqnarray}
(\openone\otimes T_*)(S)S&=& \sum_i (\openone\otimes A_i^\dagger)S(\openone\otimes A_i)S
\nonumber\\
&=&\sum_i (\openone\otimes A_i^\dagger)
(A^\dagger_i\otimes\openone)\nonumber\\
&=&\sum_i A_i\otimes A_i^\dagger.\nonumber
\end{eqnarray}

(iii) 
\begin{eqnarray}
 \overline{F(T,|\psi\rangle)}^{\psi} &=& \int_\psi {\rm Tr}\left[ |\psi\rangle\langle\psi^{\otimes\,2} \Omega_1(T)
\right] \nonumber\\
&=& {\rm Tr}\left[ 1/d(d+1)(\openone +S)\Omega_1(T)\right]\nonumber\\
&=& \frac{{\rm tr} \left[\Omega_1(T)+ (\openone\otimes T_*)(S)\right]}{d(d+1)}.
\nonumber
\end{eqnarray}

Notice that the second term inside the square brackets is, for unital
maps, simply ${\rm Tr}S=d$.  

 \end{proof}
 
Is important to stress that $\Omega_1(T)$ defined above is in general
{\em non}-Hermitian. On the other hand, $\Omega_1(T)S=(\openone\otimes T_*)(S)$ {\em is} Hermitian
(image of an Hermitian operator via  CP-map) and has the same expectation values as $\Omega_1(T)$
over symmetric states in ${\cal H}^{\otimes\,2}$. We thus associate to
$T$ a
second channel Hamiltonian:

\noindent {\bf{Definition 8}} The channel fidelity-Hamiltonian is:
\begin{equation}
  \Omega^\prime(T):=(\openone\otimes T_*)(S).
  \end{equation}

We now report, as corollaries of point (iii) above, the average
pure state fidelities of a few relevant channels.

\noindent {\bf{Corollary 4}}

(i) Mixing of unitaries from the Pauli group ($N$ qubits):
$1/(2^N+1)(1+2^N p_0)$.

(ii)
Mixing of general unitaries:
$[d+\sum_i p_i|{\rm tr}U_i|^2]/d(d+1))$.

(iii)
Amplitude damping:
$1-p(1-1/d)$.

(iv)
Projective measurements:
$[d +\sum_i |{\rm Tr} \Pi_i|^2]/d(d+1)$.

As in the channel-purity case,
the result (\ref{av_fid}) can be simply stated by saying
that: {\em the Haar-average fidelity  of a channel} $T$ {\em is given by the expectation value
of} $\Omega_1(T)$ {\em over the maximally mixed state} $\Pi_+({\cal
  H})= (\openone +S)/d(d+1)$ {\em over the symmetric subspace of} ${\cal H}^{\otimes\,2}$.
We note that a formula related to Eq.~(\ref{av_fid}) for the average fidelity of quantum operations
has been given in Ref.~\cite{man}. 

\section{Conclusions and Outlook}
\label{Conclusions}

We have introduced a Hamiltonian operator formalism for the calculation of the
channel purity and pure state fidelity. Using this formalism we have
been able to analytically compute these measures for a variety of
channels of interest in the theory of open quantum systems, and
quantum information theory. These analytical results are restricted to
cases where the eigenstates of the Hamiltonian $\Omega$ (or
$\Omega^\prime$) are product states in the symmetric subspace of ${\cal
  H}^{\otimes 2}$. When this is not the case one may have to resort to
numerical methods to compute the purity/fidelity.

A tempting generalization of our method is to consider perturbations
to the channel Hamiltonian and use the well-developed tools
of perturbation theory to thus study perturbations to given
channels. One may further speculate about an adiabatic approximation,
wherein slowly time-dependent channels can be studied using the
adiabatic theorem applied to the channel Hamiltonian. We leave these
as subjects for future investigations.

\acknowledgments
 P.Z.  gratefully acknowledges financial support by
Cambridge-MIT Institute Limited and by the European Union project  TOPQIP
(Contract IST-2001-39215).. D.A.L. gratefully acknowledges financial support from
NSERC and the Sloan Foundation.


\begin{thebibliography}{10}

\bibitem{Breuer:book}
{H.-P. Breuer and F. Petruccione}, {\em {The Theory of Open Quantum Systems}}
  ({Oxford University Press}, Oxford, 2002).

\bibitem{Nielsen:book}
{M.A. Nielsen and I.L. Chuang}, {\em {Quantum Computation and Quantum
  Information}} ({Cambridge University Press}, Cambridge, UK, 2000).

\bibitem{Brumer:book}
{P.W. Brumer and M. Shapiro}, {\em {Principles of the Quantum Control of
  Molecular Processes}} ({Wiley}, 2003).

\bibitem{Peres:book}
A. Peres, {\em Quantum Theory: Concepts and Methods} (Kluwer, Dordrecht, 1998).

\bibitem{Kraus:83}
{K. Kraus}, {\em {States, Effects and Operations}}, {\em {Fundamental Notions
  of Quantum Theory}} ({Academic}, Berlin, 1983).

\bibitem{Shor:95}
{P.W. Shor}, Phys. Rev. A {\bf 52},  2493  (1995); {A.M Steane},
Phys. Rev. Lett. {\bf 77}, 793 (1997).

\bibitem{Knill:97b}
{E. Knill and R. Laflamme}, Phys. Rev. A {\bf 55},  900  (1997).

\bibitem{Gottesman:97}
{D. Gottesman}, Phys. Rev. A {\bf 54},  1862  (1996).

\bibitem{Steane:99}
For a review see {A.M. Steane},  in {\em {Introduction to Quantum Computation and Information}},
  edited by {H.K. Lo, S. Popescu and T.P. Spiller} ({World Scientific},
  Singapore, 1999), pp. {184--212}.

\bibitem{Zanardi:97a} {P. Zanardi and M. Rasetti}, Phys. Rev. Lett.  {\bf 79}, 3306 (1997);
Mod. Phys. Lett. B {\bf 11},  1085  (1997).

\bibitem{LidarWhaley:03}
{D.A. Lidar, I.L. Chuang, and K.B. Whaley}, Phys. Rev. Lett.  {\bf
  81}, 2594 (1998); For a review see {D.A. Lidar and K.B. Whaley}, in {\em {Irreversible Quantum Dynamics}}, Vol.~622 of {\em Springer Lecture
  Notes in Physics}, edited by {F. Benatti and R. Floreanini} (Springer,
  Berlin, 2003), p. 83, eprint quant-ph/0301032.

\bibitem{Shor:03}
{P.W. Shor}, quant-ph/0305035, to appear in {Comm. in Math. Phys.}.

\bibitem{Zanardi:00}
{P. Zanardi, C. Zalka, and L. Faoro}, Phys. Rev. A {\bf 62},  {030301(R)}
  (2000).

\bibitem{Lidar:PRL99}
{D.A. Lidar, D. Bacon and K.B. Whaley}, Phys. Rev. Lett. {\bf 82},  4556
  (1999).

\bibitem{proof} Here is a proof: ${\rm Tr} (S A\otimes B)=\sum_{ij,lm} S_{ij,lm} A_{l,i} B_{m,j}
=\sum_{ij,lm} \delta_{i,m}\delta_{j,l} A_{l,i} B_{m,j}=\sum_{ij}
A_{j,i} B_{i,j}={\rm Tr} AB$.

\bibitem{Cornwell:84I+II}
{J.F. Cornwell}, {\em {Group Theory in Physics}}, Vol.~I and II of {\em
  {Techniques of Physics: 7}} ({Academic Press}, London, 1984).

\bibitem{mapa} C. Macchiavello and G.M. Palma, Phys. Rev. A {\bf 65}, 050301 (2002).

\bibitem{man} M. Nielsen, Phys. Lett. A {\bf 303}, 249 (2002). 

\end{thebibliography}
\end{document}